\documentclass[manuscript]{aastex}
\usepackage{natbib}
\usepackage{lscape}
\usepackage{color}
\usepackage{CJKutf8}

\shorttitle{GECAM detection of a bright type-I X-ray burst from 4U~0614+09}

\begin{document}

\title{GECAM detection of a bright type-I X-ray burst from 4U~0614+09: confirmation its spin frequency at 413 Hz}

\author{Y. P. Chen$^{1}$, J. Li$^{2,3}$, S. L. Xiong $^{1}$, L. Ji$^{4}$, S. Zhang$^{1}$, W. X. Peng$^{1}$, R. Qiao$^{1}$, X. Q. Li$^{1}$, X. Y. Wen$^{1}$, L. M. Song$^{1}$, S. J. Zheng$^{1}$, X. Y. Song$^{1}$, X. Y. Zhao$^{1}$, Y. Huang$^{1}$, F. J. Lu$^{1}$, S. N. Zhang$^{1}$, S. Xiao$^{1,5}$, C. Cai$^{1,5}$, B. X. Zhang$^{1}$, Z. H. An$^{1}$, C. Chen$^{1,5}$, G. Chen$^{1}$, W. Chen$^{1}$, G. Q. Dai$^{1}$, Y. Q. Du$^{1,8}$, M. Gao$^{1}$, K. Gong$^{1}$, D. Y. Guo$^{1}$, Z. W. Guo$^{1, 12}$, J. J. He$^{1}$, B. Li$^{1}$, C. Li$^{1,7}$, C. Y. Li$^{1,6}$, G. Li$^{1}$, J. H. Li$^{1,5}$, L. Li$^{1}$, Q. X. Li$^{1,7}$, X. B. Li$^{1}$, Y. G. Li$^{1}$, J. Liang$^{1,8}$, X. H. Liang$^{1}$, J. Y. Liao$^{1}$, J. C. Liu$^{1}$, X. J. Liu$^{1}$, Y. Q. Liu$^{1}$, Q. Luo$^{1,5}$, X. Ma$^{1}$, B. Meng$^{1}$, G. Ou$^{1}$, D. L. Shi$^{1,8}$, F. Shi$^{1}$, J. Y. Shi$^{1}$, G. X. Sun$^{1}$, X. L. Sun$^{1}$, Y. L. Tuo$^{1}$, C. W. Wang$^{1}$, H. Wang$^{1}$, H. Y. Wang$^{1}$, J. Wang$^{1}$, J. Z. Wang$^{1}$, P. Wang$^{1}$, Y. S. Wang$^{1}$, Y. X. Wang$^{1}$, X. Wen$^{1}$, H. Wu$^{1,8}$, S. L. Xie$^{1,13}$,, Y. B. Xu$^{1}$, Y. P. Xu$^{1}$, W. C. Xue$^{1}$, S. Yang$^{1}$, M. Yao$^{1,8}$, J. Y. Ye$^{1}$, Q. B. Yi$^{1,7}$, C. M. Zhang$^{1}$, C. Y. Zhang$^{1,9}$, D. L. Zhang$^{1}$, Fan Zhang$^{1}$, Fei Zhang$^{1}$, H. M. Zhang$^{1}$, K. Zhang$^{1, 10}$, P. Zhang$^{1,8}$, X. L. Zhang$^{1,10}$, Y. Q. Zhang$^{1,5}$, Z. Zhang$^{1}$, G. Y. Zhao$^{1,7}$, S. Y. Zhao$^{1,8}$, Y. Zhao$^{1,11}$, C. Zheng$^{1,5}$, X. Zhou$^{1,5}$, Y. Zhu$^{1}$}
 
\email{chenyp@ihep.ac.cn, jianli@ustc.ac.cn,xiongsl@ihep.ac.cn,jilong@mail.sysu.edu.cn}

\affil{$^{1}$ Key Laboratory for Particle Astrophysics, Institute of High Energy Physics, Chinese Academy of Sciences, 19B Yuquan Road, Beijing 100049, China}
\affil{$^{2}$ CAS Key Laboratory for Research in Galaxies and Cosmology, Department of Astronomy, University of Science and Technology of China, Hefei 230026, China}
\affil{$^{3}$ School of Astronomy and Space Science, University of Science and Technology of China, Hefei 230026, China}
\affil{$^{4}$ School of Physics and Astronomy, Sun Yat-Sen University, Zhuhai, 519082, China}
\affil{$^{5}$ University of Chinese Academy of Sciences, Chinese Academy of Sciences, Beijing 100049, China}
\affil{$^{6}$ Physics and Space Science College, China West Normal University, Nanchong 637002, China}
\affil{$^{7}$ Key Laboratory of Stellar and Interstellar Physics and Department of Physics, Xiangtan University, 411105 Xiangtan, Hunan Province, China}
\affil{$^{8}$ School of Computing and Artificial Intelligence, Southwest Jiaotong University, Chengdu, 611756, China}
\affil{$^{9}$ Changchun University of Science and Technology, Changchun 130022, Jilin, China}
\affil{$^{10}$ College of Physics and Engineering, Qufu Normal University, Qufu 273165, China}
\affil{$^{11}$ Department of Astronomy, Beijing Normal University, Beijing 100875, China}
\affil{$^{12}$ College of physics Sciences \& Technology, Hebei University, Baoding City, Hebei Province 071002, China}
\affil{$^{13}$ Institute of Astrophysics, Central China Normal University, Wuhan 430079, China}

\begin{abstract}

One month after launching Gravitational wave high-energy Electromagnetic Counterpart All-sky Monitor (GECAM), a bright thermonuclear X-ray burst from 4U~0614+09
was observed on January 24, 2021. 
We report the time-resolved spectroscopy of the burst and a burst oscillation detection at 413 Hz with a fractional amplitude   3.4\% (rms). This coincides with the burst oscillation previously discovered with \textit{Swift}/BAT \citep{Strohmayer2008}, and therefore confirms the spin frequency of this source.
This burst is the brightest one in the normal bursts (except the superburst) ever detected from 4U~0614+09, which leads to an upper limit of distance estimation as 3.1 kpc.
The folded light curve during the burst oscillation shows a multi-peak structure, which is the first case observed during a single burst oscillation  in nonpulsating sources.
The multi-peak profile could be due to   additional harmonics of the burst oscillation,
which is corresponding to several brighter/fainter spots at the stellar surface. 

%


\end{abstract}

\section{Introduction} 

Type I X-ray bursts (also known as thermonuclear bursts, “bursts” hereafter) were discovered in the mid-1970s \citep{Grindlay,Belian}. 
They are due to unstable burning of the accreted matter in the atmosphere of neutron stars supplied by the companion
star in low-mass X-ray binaries (LMXBs) (for reviews,  see \citealp{Lewin,Cumming,Strohmayer2006,Galloway}).
They exhibit spikes in X-ray lightcurves with a fast rise followed by an exponential decay, with a time-scale of 10--100 seconds.  
Their spectra show a blackbody profile with an evolving temperature of $kT\sim$1–-3 keV.

Some strong bursts often exhibit a plateau in the luminosity diagram and an  enhancement of the blackbody emission area. It is proposed that the energetic burst emission, i.e., reaching the Eddington limit, temporarily lifts the neutron star photosphere  due to radiation pressure. Under this assumption, with a known distance, the mass and radius of neutron star could be inferred.

Nearly 20\% bursts show oscillations with fractional amplitudes in the range of 2\%--20\% rms.
The oscillation periods are believed to be around their spin periods although drifting by several Hz in some cases (for reviews,  see \citealp{Watts2012,Bilous2019}).
Among 115 galactic X-ray bursters, there are only 19 sources in which  the detections of burst oscillations are considered robust\footnote{https://staff.fnwi.uva.nl/a.l.watts/bosc/bosc.html}. 
 Among 11 out of the 19 sources, the pulse signal is absent in the persistent/accretion emission, i.e., they are not 
		accretion-powered millisecond X-ray pulsar. 
To explain the origin of thermonuclear burst oscillations (TBOs), a brightness asymmetry on the neutron surface is required, which could be driven by different mechanisms, e.g.,   asymmetries  ignition  (hotspots) across the neutron stars surface \citep{Strohmayer1998, Strohmayer1999},   excitation  of  large-scale and low-frequency waves (r modes) \citep{Heyl2004,Chambers2019} in the neutron star ocean and   spreading of a cooling wake to form vortices in the stellar surface    \citep{Mahmoodifar2016}.
The former and the latter two models are thought to be related to the TBOs appearance in the rising and decaying phase of the bursts.
%
However, all these mechanisms fail to explain certain aspects of observations, including the absence of TBO during the majority of bursts, up to 48\% fractional amplitude of TBOs \citep{Mahmoodifar2019} and the magnetic field affection on the TBO detection.
%

The TBO signal of 4U 0614+09 at 415\,Hz was first observed with \textit{Swift}/BAT in the tail of a burst \citep{Strohmayer2008},  with a  4 $\sigma$ significance assuming a conservative number of trials. 
However, this TBO was only seen in one time window from one burst, which does not qualify 4U 0614+09 as a robust TBO source \citep{Bilous2019}.

4U 0614+09 is a faint burster and a persistent LMXB located at the anti-galactic-center direction with a distance of $\sim$ 3.2 kpc and a luminosity mostly dithering $\sim$ 1\% $L_{\rm Edd}$ \citep{Kuulkers2010}.
Although 4U 0614+09 is one of the closest bursters, only a small number of bursts have been recorded since its discovery in the 1970s. This is due to the long burst recurrence time (on a time scale of weeks) caused by the low accretion rate \citep{Linares2012}.
Most of these burst samples were observed by telescopes with a large field of view, e.g., Fermi/GBM and HETE-2/FREGATE.


Thanks to the wide field of view and the high sensitivity to X-ray photons down to about 6 keV, the Gravitational wave high-energy Electromagnetic Counterpart All-sky Monitor (GECAM) \citep{Zhang2019}  detected a burst from 4U 0614+09 during its commissioning phase.  
We report the   spectral and timing results of this   thermonuclear burst from 4U 0614+09 in Section 2 and  Section 3.
In the last Section, the discussion is given.



\section{Observations and Data analysis}

GECAM is a pair of X-ray and gamma-ray all-sky monitors (i.e. GECAM-A and GECAM-B), which aims to search for gamma-ray counterparts to gravitational wave events, which was launched from China Xichang Satellite Launch Center on December 10, 2020 (Beijing time). Two GECAM satellites share the same orbit with an altitude of 600 km and inclination angle of 29 degrees. 

Each GECAM satellite has 25  gamma-ray Detectors (GRDs) and 8 charged particle detectors (CPDs).
A GRD have a detector diameter 3 inch in diameter, works in 6 keV–5 MeV with  a time resolution down to 0.1 us and a energy resolution of 5.3\% at 662 keV, which  is comprised of LaBr$_{3}$:Ce crystal  and the SiPM array.

Each GECAM satellite is designed to cover more than a half sky (i.e. unocculted region by the Earth), thus two satellites in principle could monitor the whole sky. However, there are unexpected power supply issues on satellites, detectors on GECAM-A has not been turned on yet, thus only GECAM-B is used in the rest of this paper.

GECAM-B was triggered in-flight by a long burst at 2021-01-24T11:50:03.600 UTC (MJD 59238.49309722, denoted as $T_{0}$) \citep{Xiong2021}, which is also used as a reference to extract lightcurve and spectrum.
Within the detection location error, 4U~0614+091 lies 2.6$^{\circ}$  away from the in-flight location. 
Based on the spectral evolution and in particular the burst oscillation detection as shown below, we confirm that this burst is a genuine thermonuclear X-ray burst from 4U 0614+09 \citep{Chen2021}.

\begin{figure}[t]
\centering
   \includegraphics[angle=0, scale=0.6]{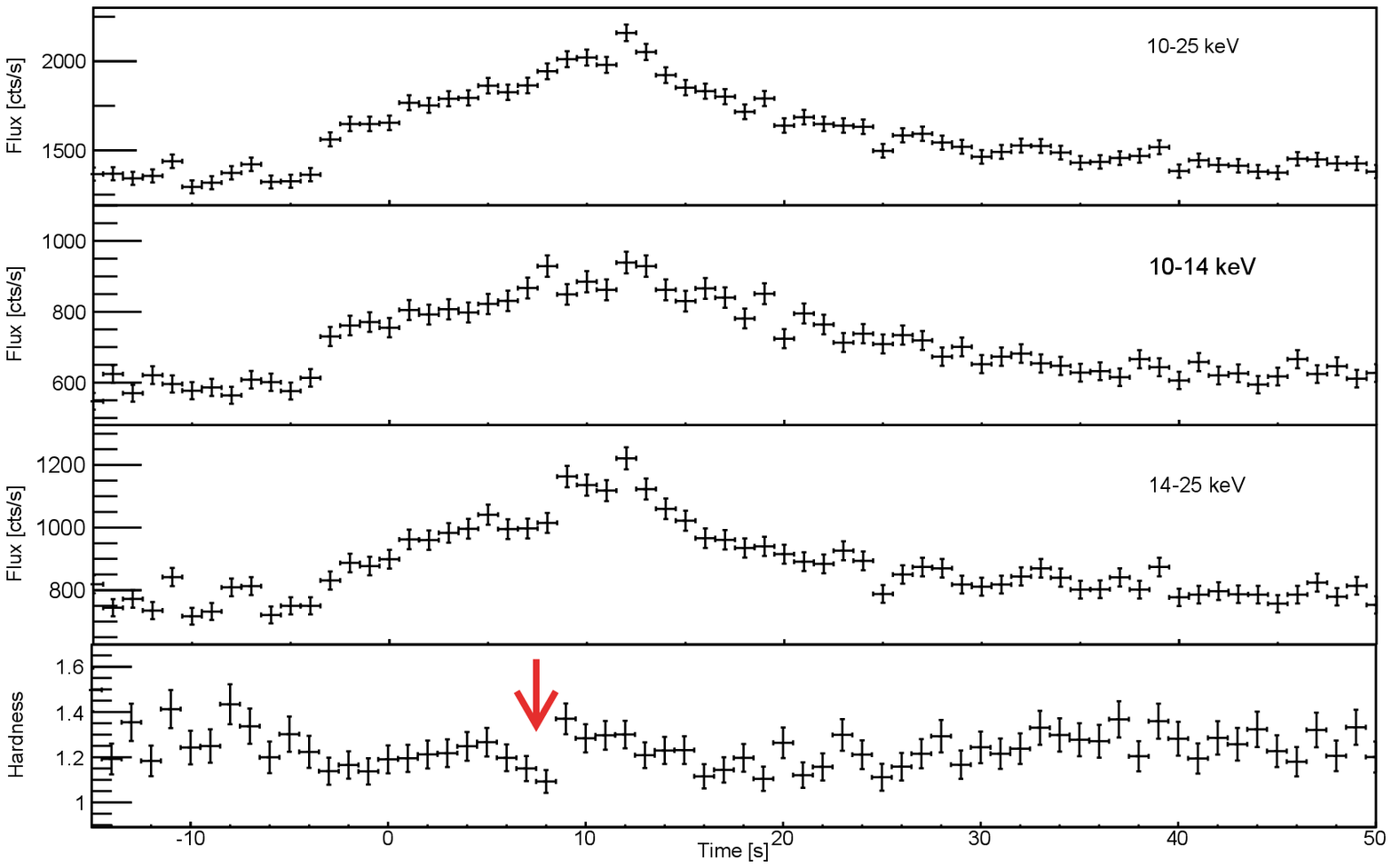}
 \caption{The burst light curves in 10--25 keV, 10--14 keV, 14--25 keV and the harness ratio of the flux in 14--25 keV to 10--14 keV. The red arrow in the bottom panel marks the dip in the hardness, which indicates a possible PRE during the burst.
   }
\label{lc_burst}
\end{figure}

Lightcurves of all the 25 GRDs are extracted and examined one by one according  to whether there is a spike around  $T_{0}$, 
among which we find the lightcurvs from detectors \#11, \#12, \#21 and \#22 have spikes around $T_{0}$.
Therefore, data from these four detectors are used in the following analysis.
The lightcurves in 10--25 keV, 10--14 keV, 14-25 keV and
the hardness ratio of the   count rates  
of 10--14 keV to 14--25 keV are shown in Fig. \ref{lc_burst}. Beyond this energy band, the count rates are dominated by the background emission.

We performed a time-resolved spectroscopy of the burst using a 2\,s step. The pre-brust emission was subtracted off as the background during the burst spectral fitting. 
Response files were generated using {\sc RSP\_Generator} from GECAM CALDB v0.3\footnote{http://gecamweb.ihep.ac.cn/xgwd.jhtml}.
The time-resolved spectra were fitted using an absorbed blackbody  model in the energy range of 10--25 keV, assuming  a "TBABS" model \citep{Wilms2000} and an interstellar hydrogen column density  $N_{\rm H}$ fixed at 0.346 $\times10^{-22}~{\rm cm}^{-2}$ \citep{Ludlam2019}.
The software package {\sc Xspec v12.10.0} was used for the spectral fitting.
For the distance  in our all calculation,  including the luminosity and blackbody radius,  we use for a simplicity value of 3 kpc, since the distance is estimated under 15\% uncertainty \citep{Kuulkers2010}.




\section{Results}

\subsection{Burst spectral evolution}
As shown in Fig. \ref{lc_burst}, the burst has a rise time of 10\,s followed by an exponential decay, with a peak flux of 700 cts/s higher than the pre-burst emission in the energy range of 10--25\,keV.
The hardness diagram indicates a spectral softening during its decay phase.
The forementioned profile is common across normal type-I X-ray bursts.  

\begin{figure}[t]
\centering
   \includegraphics[angle=0, scale=0.4]{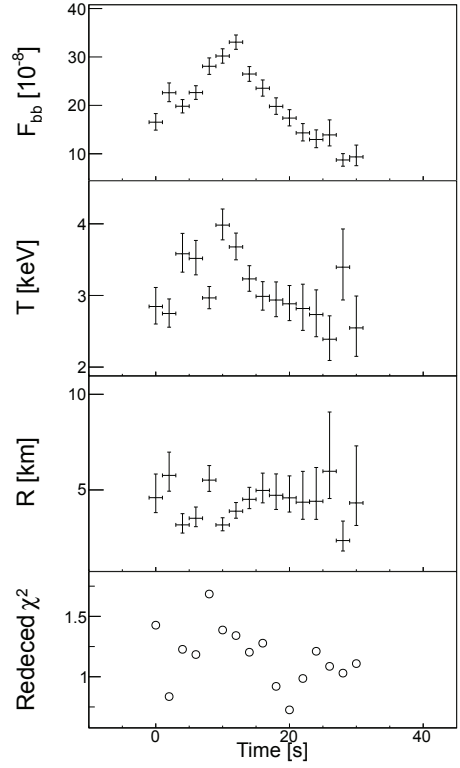} 
 \caption{Time-resolved spectroscopy of the X-ray burst using
an absorbed  blackbody model, with a time step 2\,s.
The unabsorbed bolometric flux $F_{\rm bb}$ in a unit of $10^{-8}~{\rm erg}~ {\rm cm}^{-2}~{\rm s}^{-1}$ (top panel), the  temperature $kT_{\rm }$ (2nd panel), the radius $R$ at a distance of 3 kpc (3rd panel) and the reduced $\chi^{2}$ statistic (bottom bottem) are given in the 4 panels, respectively.
}
\label{burst_evolution}
\end{figure}

\begin{figure}[t]
\centering
   \includegraphics[angle=0, scale=0.5]{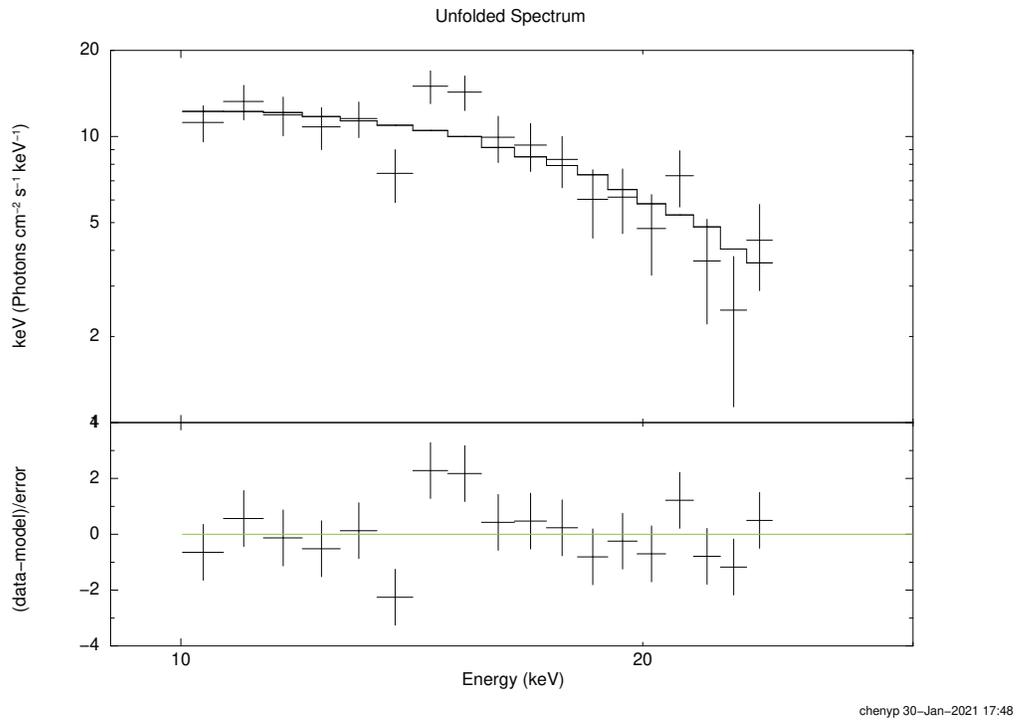}
 \caption{The spectra of the burst peak emission fitting by a blackbody model. 
 }
\label{spe_burst}
\end{figure}

 As shown in Fig. \ref{burst_evolution}, the best-fitting parameters are the blackbody temperature $T_{\rm bb}$ $\sim$ 2--4 keV, the radius $R_{\rm bb}$ $\sim$ 6\,km   with a reduced $\chi_{\upsilon}$  $\sim$0.7--1.7 (16 degrees of freedom).  The spectral fitting results and residuals  at the burst peak time are shown in Fig. \ref{spe_burst}. 
The temperature has a substantial decay along the burst decay phase, which is characteristic to thermonuclear bursts.
 The burst  bolometric peak flux is $\sim 33.0 \pm 1.5$$\times 10^{-8}~{\rm erg}~{\rm cm}^{-2}~{\rm s}^{-1}$, 
which is corresponding to  $\sim3.56\pm 0.2$$\times10^{38}$ erg s$^{-1}$ at a distance of 3.0 kpc.
We notice that the peak flux is  brighter than the previously report  bursts   ($\sim 31.5 \pm 0.5$$\times 10^{-8}~{\rm erg}~{\rm cm}^{-2}~{\rm s}^{-1}$; \citealp{Kuulkers2010}) of 4U~0614+91 except for the super-burst, casting some doubt on whether this burst have a PRE phase.
 
 

The burst fluence integrating the unabsorbed bolometric luminosities over the whole burst is 7.5$\pm$0.3 $\times 10^{-6}~{\rm erg}~{\rm cm}^{-2}$, which corresponds to  8.1$\pm$0.3 $\times 10^{39}~{\rm erg}$ assuming a distance of 3 kpc.
The burst duration ($\tau$=$E_{\rm b}/F_{\rm pk}$, ration of the fluence to the peak flux) is 22.7$\pm$1.4 s.

The above analysis does not include the gravitational redshift and spectral hardness correction due to the scattering of the photosphere, i.e., the parameters above are apparent values observed from a distant observer.
This correction will be given in the discussion part of this work.

\subsection{Burst oscillation}

\begin{center}
\begin{figure*}
\centering
\includegraphics[scale=0.42]{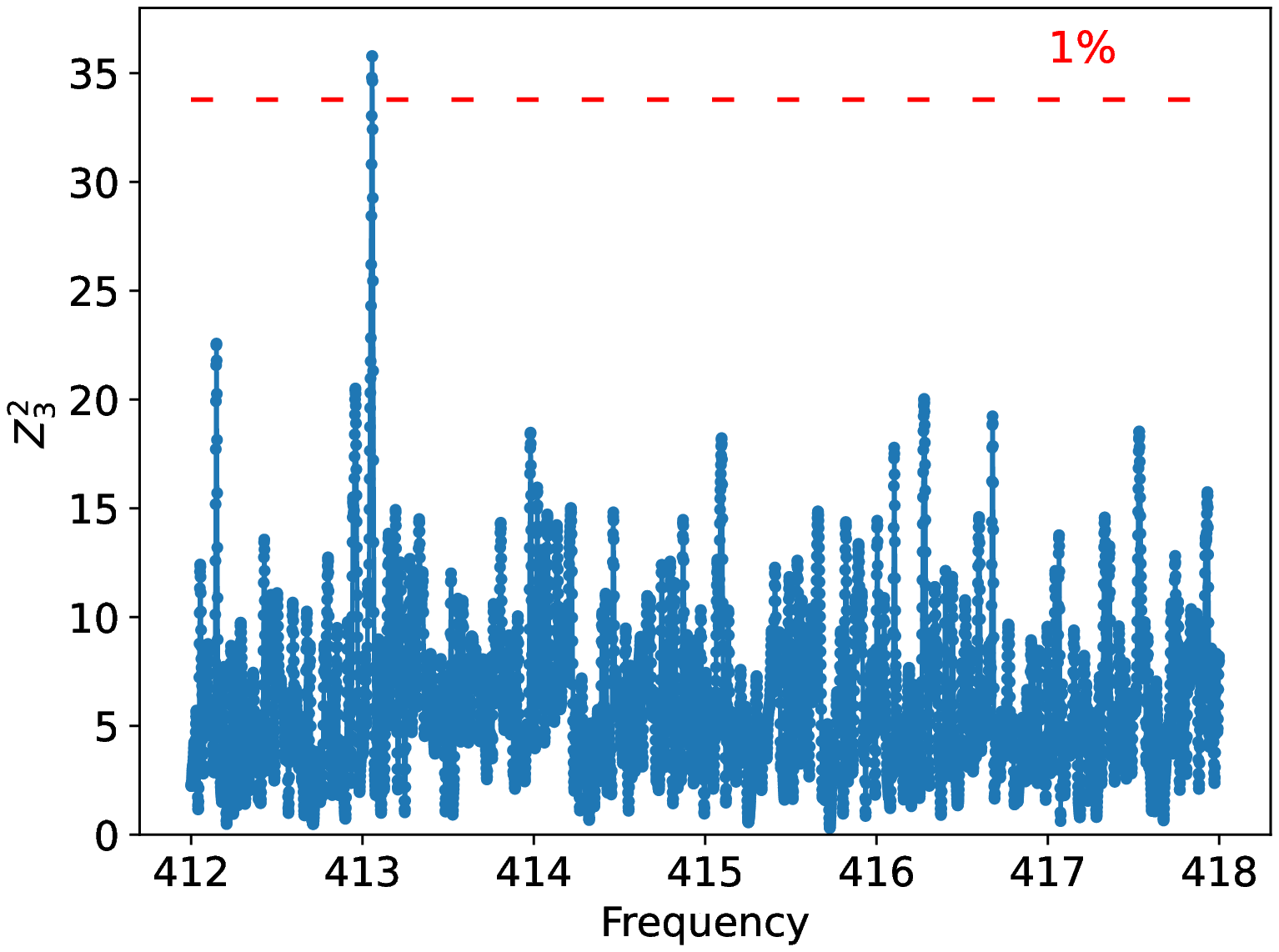}
\includegraphics[scale=0.4]{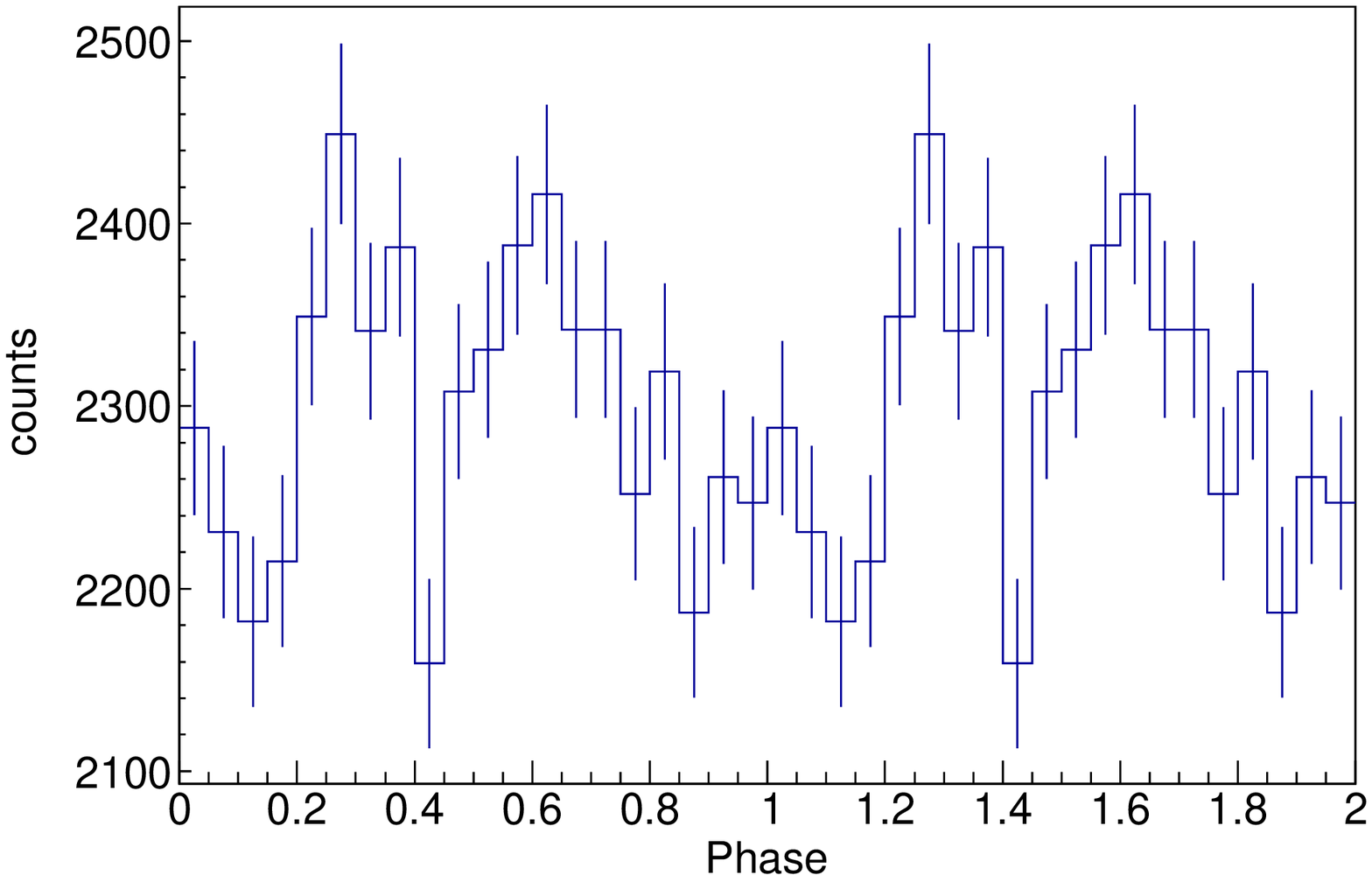}
\includegraphics[scale=0.6]{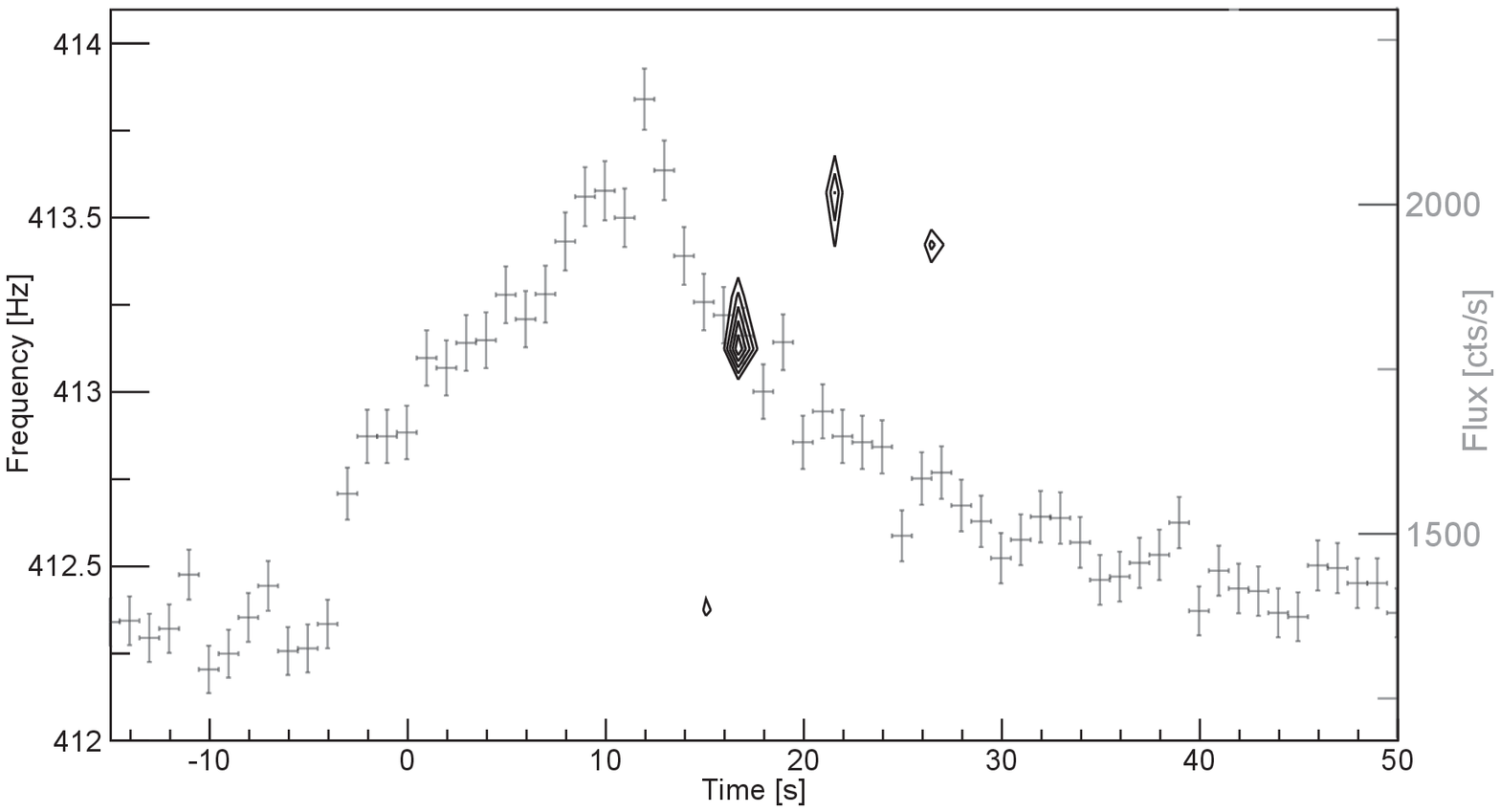}
\caption{Left: Z$^{2}_{3}$ periodogram of 4U~0614+09 during the type-I burst. The red dashed line indicates false alarm probability of 1\% level.
Right: Phase light curve of 4U~0614+09 with the best detected frequency.
%
Bottom: The Power spectra using 2 s intervals and stepped the intervals by 0.5 s, five contour levels of the Z$^{2}_{3}$ are plotted, starting at 15 and spaced in steps of 2.
}
\label{Z2}
\end{figure*}
\end{center}

We considered the burst data within $T$=0--30 s for timing analysis.
Pulsations were searched via an Z$^{2}_{n}$-test procedure in a narrow frequency range (412--418 Hz) around previously known frequency ($\sim 415$ Hz, \citealp{Strohmayer2008}), with a step of 0.003 Hz and a number of harmonics $n$ varied from 1 to 4.
We start from Z$^{2}_{1}$-test and {increase $n$ until a significant} signal is detected.
Since no burst activity was observed above 25 keV, only photons in 10--25 keV were included in the timing analysis.
We found a peak at $F=413.056$ Hz with a Z$^{2}_{3}$ statistic of 35.54 (Figure \ref{Z2}, left panel), which corresponds to a single frequency significance of 4.65 $\sigma$.
To estimate the chance that the timing signal detected above is originated from noise, we calculated the false alarm probability via implementing a bootstrap method.
A simulated event list was constructed. 
We randomly scattered the same number of observed events in the time range of the burst and carried out the Z$^{2}_{3}$-test on 10$^{4}$ simulated event lists and derived the false alarm probability of the detected timing signal, which is 6$\times$10$^{-3}$.
All the sampled frequencies in the Z$^{2}_{3}$-test have been considered in the bootstrap.
Our results are shown in Figure \ref{Z2}, left panel.

We folded the burst events using the best detected frequency (Figure \ref{Z2}, right panel).
Based on the folded light curve, the burst oscillation fractional rms amplitude  is 3.4\%.
 The phase light curve shows a multi-peak structure, which is different from the sinusoidal profile observed previously in 4U 0614+09 with \textit{Swift}/BAT \citep{Strohmayer2008}.
It is quite uncommon and the first time observed among burst oscillations.
It may arise from frequency changes/drifts during the type-I burst.
To explore this possibility, we searched for timing signals using a sliding window technique, with a data window of 2 s and a sliding step of 0.5 s.
We run Z$^{2}_{3}$ timing analysis with a frequency resolution of 0.01Hz in each data window, and constructed a 2D plane (Figure \ref{Z2}, bottom panel). 
The timing signal was only significantly detected in a few data windows.
The most significant detection is in data window 16--18 s, at $F=413.056$ Hz with a Z$^{2}_{3}$ statistic of 30.01 (Figure \ref{Z2}, bottom panel), which corresponds to a single frequency significance of 4.1 $\sigma$.
Simulations were carried out to estimate the false alarm probability, leading to a value of 2.7$\times$10$^{-3}$.
The phase light curve within this data window, or other less significant hot spots in Figure \ref{Z2}, bottom panel, also shows a multi-peak profile.
The unique profile of burst oscillation observed in this type-I burst of 4U~0614+09 might be authentic, but could not be further tested because of limited statistics.

\section{Discussion}

We notice that the observed blackbody radius increases as the flux decrease along with burst cooling, which is not consistent with the emission region corresponding to the whole stellar surface.
A possible explanation for this inconformity 
is related to a spectral hardening factor (color factor) due to electron scattering occurs at the atmosphere outlayer  \citep{London1986}.
This spectral harden factor ($f_{\rm c}\equiv T_{\infty}/T_{\rm eff}$) depends only on the burst luminosity for a hydrogen-poor composition. It makes the emergent emission deviate from Planck curve, manifesting an overestimation of the temperature by a factor of $f_{\rm c}$  and an underestimation of the radius by a factor  of 1/$f_{\rm c}^{2}$. 
For  bursts with luminosities brighter than 0.2 $L_{\rm Edd}$, $f_{\rm c}$ lies between 1.4 and 1.7, and increases with burst luminosity $L_{\rm bb}$, which causes the  anti-correlation between the radius and the flux.

Recent observations found that the stellar atmosphere model also  depends on the accretion rate and spectral state of the source. For instance,
in atoll sources,  the stellar atmosphere model (the spectral hardening factor of bursts) is no longer functional in the high/soft state (the banana state) \citep{Suleimanov2011,Kajava2014,Suleimanov2018}.
However, in our analysis it is difficult to constrain the accretion-state  based on GECAM observation, or \textit{Swift}/BAT and MAXI daily lightcurves, though it appears that 4U 0614+09 spent most of its time in low/hard state (the island state).

We also notice the peak temperature  reached up to $4.0\pm0.2$ keV, which is a relatively high value for burst temperature evolution.
Given the neutron star mass $M_{\rm NS} = 1.4M_{\odot} $ and radius  $R_{\rm NS}$ = 10 km,   a gravitational redshift factor (1 + $z$) is estimated as 1.31. 
Along with $f_{\rm c}=1.7$, the emergent temperature  observed at the stellar surface  is revised down to $T_{\infty}(1 + $z$)/f_{\rm c}=3.0\pm0.2$ keV.


Considering the detection of same frequency in two bursts of 4U~0614+09, it qualifies as the 20th source in the sample of robust burst oscillation detection. 
A dichotomy of burst oscillations is slow oscillation ($<$ 400 Hz) and rapid oscillation  ($>$ 400 Hz) \citep{Galloway,Watts2012}. 
Mostly, the former group  tends to behave short duration ($\tau<10$ s), most likely helium-dominated bursts; the latter group   tends to behave longer duration ($\tau>10$ s) that probably involves mixed hydrogen/helium burning.
A possible interpretation is that the slow rotators are largely accreting from degenerate companions with H-deficient composition,  and the rapid rotators are non-degenerate companions with more hydrogen composition, but with the exception to this rule: XTE~J1814--338, a slow oscillator of 314 Hz but with a long burst duration.
Apparently, the burst of this work is a long burst and behaves as a rapid oscillation. Based on the above model, the composition of the burning layer is hydrogen-rich. However, the donor star  is proposed to be a degenerate companion,  i.e.,   a C/O-rich donor,  which based on the presence of C and O emission lines and absence of H or He emission line on optical spectrum \citep{Nelemans2004, Werner2006}. 
Furthermore, the ignition model needs a significant amount of H/He in the accreted material \citep{Cumming2001,Kuulkers2010}, especially the bursts durations of 4U~1614+91 are several times longer than the well-known pure-He bursts of 4U~1728--34.

The multi-peak profile of the phase lightcurve is the first time reported in a single  burst oscillation  in nonpulsating sources,
 although harmonics were reported previously in   bursting-pulsars \citep{Chakrabarty2003,Strohmayer2003} and  4U~1636--536 by stacking the rising phase of nine bursts \citep{Bhattacharyya2005}, since additional harmonic  is too weak to be detected in a single burst before this work \citep{Chakraborty2014}.
A  possible reason to cause this profile is that there are  more than one antipodal brighter or fainter regions in the NS surface.
We notice that  phase-folded light curve is beyond the canonical  profile (a constant plus a sinusoid) of burst oscillations, which may indicate complicated brightness distribution in the stellar surface. 
However, the   small fractional amplitude of the TBO and the low statistics of the data   prevent us to carry out further tests.

Comparing with the other TBOs, we  notice that this TBO  is a small-amplitude oscillation. 
In the cooling wake scenario, the behavior of TBO in the burst decay phase is related to the latitude at which the burst ignites, since the first ignition spot being the first cooling spot  \citep{Ootes2017}.
The ignites latitude could be inferred from the convexity parameter of the lightcurve profile during the burst rising phase, i.e., roughly, a cancave and convex profile is responding to  high and low ignites latitude \citep{Maurer2008}.
We notice that rising profile of this burst is a convex one, and indicates that it ignites near the equator and induces a short-lived asymmetric emission due to fast speed of the cool wake during the decay phase \citep{Zhang2016}   than the case of the high latitude ignition.
 This may be related to the small fractional rms of this TBO, since TBO detectability   also depends on the time length of the asymmetric emission \citep{Maurer2008}.



 
 Last but not the least, this burst and the associated oscillation detected by GECAM proves its capability in building a thorough/unique burst sample during GECAM's service time, thanks to its large detection area and wide field of view: bursts born in the low accretion state or even in the quiescent state of sources can be caught accompanied with the detailed resolution of temporal and spectral properties. 
 
 \acknowledgements
The GECAM (Huairou-1) mission is supported by the Strategic Priority Research Program on Space Science, the Chinese Academy of Sciences, Grant No. XDA15360000.
This work is supported by the National Key R\&D Program of China (2021YFA0718500),
the Key Research Program of Frontier Sciences, Chinese Academy of Sciences, Grant NO. QYZDB- SSW-SLH012 and the National Natural Science Foundation of China under grants  11733009, U1838201, U1838202, U1938101, U2038101,
12173038. 

\bibliographystyle{plainnat}

\clearpage

\clearpage

\end{document}